\documentclass[journal]{IEEEtran}
\ifCLASSINFOpdf
\usepackage[pdftex]{graphicx}

\graphicspath{{./pdf/}{./jpeg/}}

\DeclareGraphicsExtensions{.pdf,.jpeg,.png}
\else

\fi

\usepackage{cite}
\usepackage{amsmath,amssymb,amsfonts}
\usepackage{graphicx}
\usepackage{booktabs}
\usepackage{multirow}
\usepackage{color}
\usepackage{soul}
\usepackage{algorithm}
\usepackage{algorithmic}
\usepackage{algorithmic}
\usepackage{placeins}
\usepackage{stfloats}
\usepackage{fixltx2e}
\usepackage{lineno}
\usepackage{hyperref}

\usepackage{bm}
\usepackage{stfloats}
\DeclareMathOperator*{\argmin}{arg\,min}

\usepackage{float}
\usepackage{subfigure}

\begin{document}
	\title{\Huge Reconstructing Quantitative Cerebral Perfusion Images Directly From Measured Sinogram Data Acquired Using C-arm Cone-Beam CT}
	%Direct Cerebral Hemodynamic Functional Imaging Using C-arm Cone-Beam CT and Conditional Generative Model	
	\author{Haotian Zhao, Ruifeng Chen, Jing Yan, Juan Feng, Jun Xiang, Yang Chen, Dong Liang, Yinsheng Li
		\thanks{\emph{Corresponding author: Yinsheng Li.}}
		\thanks{Haotian Zhao (e-mail: ht.zhao1@siat.ac.cn) and Ruifeng Chen (e-mail: rf.chen1@siat.ac.cn) are with Research Center for Medical Artificial Intelligence, Shenzhen Institute of Advanced Technology, Chinese Academy of Sciences, Shenzhen, China and with Laboratory of Image Science and Technology, Key Laboratory of Computer Network and Information Integration, Southeast University, Nanjing, China.}
		\thanks{Jing Yan (jing.yan@united-imaging.com), Juan Feng (juan.feng@united-imaging.com), and Jun Xiang (jun.xiang@united-imaging.com) are with X-Ray Department, United Imaging Healthcare Limited Company, Shanghai, China.}
		\thanks{Yang Chen (e-mail: chenyang.list@seu.edu.cn) is with Key Laboratory of New Generation Artificial Intelligence Technology and Its Interdisciplinary Applications, Southeast University, Nanjing, China.}
		\thanks{Dong Liang (e-mail: dong.liang@siat.ac.cn) and Yinsheng Li (e-mail: ys.li2@siat.ac.cn) are with Research Center for Medical Artificial Intelligence and Key Laboratory of Biomedical Imaging Science and System, Shenzhen Institute of Advanced Technology, Chinese Academy of Sciences, Shenzhen, China.}
		\vspace{-3em}
	}
	
	\maketitle
	\begin{abstract} 
		To shorten the door-to-puncture time for better treating patients with acute ischemic stroke, it is highly desired to obtain quantitative cerebral perfusion images using C-arm cone-beam computed tomography (CBCT) equipped in the interventional suite. However, limited by the slow gantry rotation speed, the temporal resolution and temporal sampling density of typical C-arm CBCT are much poorer than those of multi-detector-row CT in the diagnostic imaging suite. The current quantitative perfusion imaging includes two cascaded steps, time-resolved image reconstruction and perfusion parametric estimation. For time-resolved image reconstruction, technical challenge imposed by poor temporal resolution and sampling density causes inaccurate quantification of the temporal variation of cerebral artery and tissue attenuation values. For perfusion parametric estimation, it remains a technical challenge to appropriately handcraft the regularization for better solving the perfusion parametric estimation problem. These two challenges together prevent obtaining quantitatively accurate perfusion images using C-arm CBCT. The purpose of this work is to simultaneously address these two challenges via combining the two cascaded steps into a single joint optimization problem and reconstructing quantitative perfusion images directly from the measured sinogram data. In the developed direct cerebral perfusion parametric image reconstruction technique, TRAINER in short, the quantitative perfusion images have been represented as a subject-specific conditional generative model trained under the constraint of the time-resolved CT forward model, perfusion convolutional model, and the subject's own measured sinogram data. Results shown in this paper demonstrated that using TRAINER, quantitative cerebral perfusion images can be accurately obtained using C-arm CBCT in the interventional suite. 
	\end{abstract}
	
	\begin{IEEEkeywords}
		C-arm Cone-Beam CT; Quantitative Cerebral Perfusion Imaging; Conditional Generative Model; Deep Learning; Acute Ischemic Stroke 
	\end{IEEEkeywords}
	
	\IEEEpeerreviewmaketitle
	
	%%%%%%%%%%%%%%%%%%%%%%%%%%%%%%%%%%%%%%%%%%%%%%%%%%%%%%%%%%%%%%%%%%%%%%%%%%%%%%%%%%%%%%%%%%%%%%%%%
	\section{Introduction}
	
	\subsection{Clinical Motivation for Quantitative Cerebral Perfusion Imaging Using C-arm Cone-Beam CT}
	
	Quantitative cerebral perfusion imaging aims to assess the feasibility of endovascular treatment (EVT) for each individual patient suffering from an acute ischemic stroke. In the current clinical practice, quantitative cerebral perfusion imaging is mainly implemented with either multi-detector-row computed tomography (MDCT) \cite{RN9} or magnetic resonance (MR) imaging \cite{RN126} in the diagnostic imaging suite, can not be implemented in the interventional suite. The time needed to perform these exams, waiting and transportation of patients among different suites result in a significant delay from patient's arrival to treatment. According to a recent clinical study \cite{RN65}, a 30-minutes or 60-minutes delay in the treatment causes an 11$\%$ or 38$\%$ reduction in functional outcome respectively. Therefore, shortening the door-to-puncture time leads to better functional outcome. 
	
	Although the hybrid interventional suite combining either CT-based or MR-based workflow with the C-arm CBCT may be available in some world-leading medical centers, its availability is still largely limited. Therefore, to shorten the door-to-puncture time, it is highly desired to conduct all needed imaging tasks especially the quantitative perfusion imaging directly using C-arm CBCT in the interventional suite bypassing the CT-based or MR-based workflow \cite{niu2016c,yang2015time,li2019enhanced}. Once the patient is identified to be feasible for EVT, the treatment can be executed immediately without any delay. 
	
	%quantify the volumes of reversible tissues (a.k.a. penumbra) and irreversible tissues (a.k.a. infarct core) for 
	%consisting of non-contrast CT, CT angiography, CT perfusion or MR angiography, diffusion weighted imaging, fluid attenuated inversion recovery, perfusion weighted imaging
	
	\subsection{Technical Challenges for Quantitative Cerebral Perfusion Imaging Using C-arm Cone-Beam CT}
	
	The current quantitative perfusion imaging includes two cascaded steps, time-resolved image reconstruction and perfusion parametric estimation. Compared with MDCT, temporal resolution and temporal sampling density of typical C-arm CBCT are inherently limited for quantitative perfusion imaging. For intravenous injection, to cover the entire time course of the contrast uptake, the total data acquisition time is typically about 30-60 seconds. Using MDCT, this can be obtained by continuously acquiring data at a speed as fast as 0.25 seconds per time frame. Using C-arm CBCT, this can be obtained by acquiring data with the temporal resolution (4-8 seconds per time frame) and temporal sampling density (6-10 time frames per exam). For time-resolved image reconstruction, technical challenge imposed by poor temporal resolution and poor sampling density causes inaccurate quantification of the temporal variation of cerebral tissue attenuation values and hence significantly limit quantitative accuracy of quantitative perfusion imaging using C-arm CBCT.  
	
	\subsection{Techniques to Address Technical Challenges for Time-Resolved Image Reconstruction Using C-arm Cone-Beam CT}
	
	Several techniques have previously been proposed to address the potential challenges in time-resolved image reconstruction using C-arm CBCT. One scheme \cite{neukirchen2010iterative,wagner2013model,manhart2013dynamic} was to decompose the time attenuation curves (TACs) of different image voxels in terms of known temporal basis functions such as Gaussian functions, gamma-variate functions, linear or cubic splines. The incorporation of known basis functions dramatically reduces the number of unknowns, hence, strongly regularizes the otherwise ill-posed optimization problem. In the second scheme, a temporal deconvolution method was proposed in \cite{tang2013novel} to improve the temporal resolution and temporal sampling density by deconvolving the reconstructed TACs with a known convolution kernel simulating the temporal resolution degradation scenario. In the third scheme, a local TACs optimization framework was proposed in \cite{van2016local} to model the TACs inside the vessel regions and optimize the TACs constrained by the acquired data. The fourth scheme was to improve temporal resolution via the limited-view image reconstruction strategy. One such method is the SMART-RECON \cite{chen2015synchronized}, which reconstructs up to four time frames with temporal variation without suffering from severe limited-view artifacts using data acquired over the short-scan range \cite{li2018time}. By incorporating the prior knowledge of periodicity of the limited-view artifacts in the C-arm CBCT perfusion data acquisition scheme, eSMART-RECON \cite{li2019enhanced} further improves the temporal resolution to 4-7.5 frames per second (fps). Additionally, by recasting the temporal resolution improvement task as a temporal extrapolation task, AIRPORT \cite{Li2023airport} improves the temporal resolution to 40 fps, the time window needed to acquire a single projection view data using typical C-arm CBCT and the data acquired over the short-scan range. 
	
	\subsection{Technical Challenges for Perfusion Parametric Estimation}
	
	Perfusion parametric estimation aims to quantify cerebral blood flow (CBF), cerebral blood volume (CBV), mean transit time (MTT), and time to peak (TTP) etc using the reconstructed time-resolved CT images. For perfusion parametric estimation, currently available techniques \cite{RN74,RN73,RN72,RN71,RN70,RN69,RN68} incorporate handcrafted regularization to heavily constrain the otherwise ill-posed deconvolution problem especially under non-ideal data acquisition conditions such as low exposure levels. However, it remains a technical challenge to appropriately design the regularization for better solving the deconvolution problem. Inappropriate designs of regularizations may lead to bias in the obtained parametric images \cite{RN55,RN41}, even if the used time-resolved images are quantitatively accurate. 
	
	%These methods did not incorporate the CT data acquisition model, hence fail to leverage the measured data to constrain the deconvolution problem.
	
	\subsection{Purpose and Innovations of This Work}
	
	Although significant efforts have been made to reconstruct time-resolved images and estimate perfusion parametric images from the reconstructed time-resolved CT images, no work has yet been published to address inaccurate perfusion imaging jointly caused by the poor temporal resolution of C-arm CBCT and current perfusion parametric estimation techniques due to biased regularization. These two challenges together prevent the acquisition of quantitatively accurate perfusion images using C-arm CBCT.
	
	In the current quantitative perfusion imaging pipeline, time-resolved image reconstruction and perfusion parametric estimation are treated as two cascaded challenges. It is our conviction that addressing each of them independently may fail to achieve the needed overall accuracy. On the contrary, in this work, instead of addressing each challenge independently, we recast the quantitative perfusion imaging problem as a single joint optimization problem and reconstruct quantitative perfusion images directly from the measured sinogram data of each individual subject. The proposed single joint optimization strategy has two unique advantages. First, it significantly reduces the number of unknowns to be determined from the number of time frames of the time-resolved images to the number of perfusion parametric images, resulting in a powerful regularization to the otherwise ill-posed optimization problems. Second, the upper bound of the quantitative accuracy of perfusion images reconstructed via the proposed single joint optimization strategy only depends on the time window needed to acquire a single projection view data, not the temporal resolution of tomographic reconstruction (a.k.a., the gantry rotation speed), relaxing the requirements of C-arm rotation speed, mechanical stability or other system conditions.
	
	To implement this idea, we developed direc\underline{t} ce\underline{r}ebr\underline{a}l perfus\underline{i}o\underline{n} param\underline{e}tric image \underline{r}econstruction technique, TRAINER in short, that jointly leverages the forward model of the time-resolved CT data acquisition and the convolutional model for cerebral perfusion parametric estimation. To avoid potentially biased regularization design for perfusion parametric estimation, the perfusion parametric images in TRAINER have been represented as a subject-specific conditional generative model trained under the constraint of the subject's own measured sinogram data. 
	
	%TRAINER is free of the generalizability concern when supervised deep learning approaches are applied, i.e. reconstruction accuracy of the trained network drops for an unseen testing case. 
	
	The innovation in the proposed TRAINER is summarized as follows:
	\begin{itemize}
		\item It recasts the quantitative perfusion imaging problem as a single joint optimization problem to address poor temporal resolution of C-arm CBCT and inaccurate perfusion parametric estimation simultaneously;
		\item It explicitly leverages the measured sinogram data of each individual subject to ensure the accuracy of perfusion parametric images for each individual subject;
		\item It does not use handcrafted regularization for perfusion parametric estimation and hence avoids the error induced by inappropriate regularization.
	\end{itemize}
	
	In this work, we demonstrated that using TRAINER, quantitative perfusion images can be accurately obtained via training a subject-specific conditional generative model under the constraint of the subject's own measured sinogram data. Results shown in the paper demonstrated that the two technical challenges, i.e., poor temporal resolution of C-arm CBCT and inaccurate perfusion parametric estimation methods, have been simultaneously addressed using the proposed TRAINER technique. TRAINER may help quickly identify patients who are most likely to benefit from EVT directly using C-arm CBCT in the interventional suite.
	
	%The perfusion parametric images have been represented as a subject-specific conditional generative model and constrained by the subject's own measured data. Results shown in this paper demonstrated that using the developed XXX technique, XXX in short, cerebral perfusion parametric images can be accurately obtained using C-arm CBCT in the interventional suite. 
	
	%Since XXX learns subject-specific hemodynamic functional imaging algorithm by fitting the model parameters of XXX to each subject’s own measured data
	
	%%%%%%%%%%%%%%%%%%%%%%%%%%%%%%%%%%%%%%%%%%%%%%%%%%%%%%%%%%%%%%%%%%%%%%%%%%%%%%%%%%%%%%%%%%%%%%%%%%%%%
	\section{Direct Cerebral Perfusion Parametric Image Reconstruction}
	
	\subsection{Forward Model for Time-Resolved CT Data Acquisition}
	
	Considering the ideal CT imaging scenario where the image object is static and stationary and a monochromatic X-ray spectrum is used in the data acquisition, the forward model under proper digitization can be formulated as the following linear system:
	\begin{align}
		\label{eq:ideal-linear-system}
		\bar{\mathbf{y}} &= \mathbf{A} \mathbf{x}
	\end{align}
	where, $\bar{\mathbf{y}}\in\mathbb{R}^{M\times 1}$ denotes the modeled line integral data, $\mathbf{A}\in\mathbb{R}^{M\times N}$ denotes the CT system matrix and $\mathbf{x}\in\mathbb{R}^{N\times 1}$ denotes the unknown imaged object to be reconstructed. $M$ denotes the number of measurements associated with a single short-scan data. $N$ denotes the number of image voxels. 
	
	When the linear attenuation coefficients of the imaged object is varying, the pseudo inversion of Eq.~(\ref{eq:ideal-linear-system}) represents a temporally-averaged approximation of the true imaged object within the data acquisition time window given that the acquired data is sufficient for tomographic reconstruction. Generally, even if the imaged object is varying during the data acquisition, the image object can always be assumed to be static and stationary within the time window corresponding to a subset of continuous projection view angles.
	
	Here, we decompose the linear model in Eq.~(\ref{eq:ideal-linear-system}) to a group of linear sub-systems, each corresponding to data acquired at a subset of continuous projection view angles. The group of linear sub-systems is represented as follows:
	\begin{align}
		\label{eq:decomp-linear-systems}
		\bar{\mathbf{y}}_t &= \mathbf{A}_t \mathbf{x}_t
	\end{align}
	Where $\bar{\mathbf{y}}_t\in\mathbb{R}^{\frac{M}{K}\times 1}$, $\mathbf{x}_t\in\mathbb{R}^{N\times 1}$ denote the modeled line integral values and the image to be reconstructed associated with the $t$-th time frame respectively, $\mathbf{A}_t\in\mathbb{R}^{\frac{M}{K}\times N}$ denotes CT system matrix corresponding to $t$-th time frame, where $t=1,2,\cdots,T$ denotes the index of time frame. Here $T_0$ denotes the number of short-scan acquisitions, $K$ denotes the number of subsets within a single short-scan range, and $T=KT_0$ denotes the total number of time frames to be reconstructed.
	
	Since our goal is to jointly reconstruct the dynamic image object $\{\mathbf{x}_t\}$, it is more convenient to stack these linear sub-systems as follows:
	\begin{align}
		\label{eq:forward-model}
		\vec{\bar{\mathbf{Y}}} &= \mathcal{A} \vec{\mathbf{X}} \\ 
		\mathcal{A}&= \text{diag}\{\mathbf{A}_1, \mathbf{A}_2, \cdots, \mathbf{A}_T\} \\
		\mathbf{X} &= [\mathbf{x}_1, \mathbf{x}_2, \cdots, \mathbf{x}_T] \\
		\bar{\mathbf{Y}} &= [\bar{\mathbf{y}}_1, \bar{\mathbf{y}}_2, \cdots, \bar{\mathbf{y}}_T]
	\end{align}
	where $\mathcal{A}\in\mathbb{R}^{MT_0\times NT}$ takes each system matrix $\mathbf{A}_t$ as its diagonal matrix block, $\mathbf{X}\in\mathbb{R}^{N\times T}$ denotes the time-resolved image matrix stacking each image associated with the $t$-th time frame $\mathbf{x}_t$ along its column dimension, and $\vec{\cdot}$ denotes the vectorization operator.
	
	\subsection{Convolutional Model for Cerebral Perfusion Parametric Estimation}
	
	In patients with an AIS, the TACs may lag behind the AIF, particularly if the latter is obtained from unaffected arteries in the contralateral hemisphere \cite{miles2007multidetector}. To account for this delay, a single-compartment model is utilized to incorporate the lag effect of tissue TACs, expressing the impulse residue function as an exponential form associated with the perfusion parameter CBF and $\mathbf{\mathcal{T}}_{0}$ (time lag). $\mathbf{\mathcal{T}}_{0}$ serves as an additional valuable parameter for characterizing hemodynamic status in cerebral circulation of AIS patients \cite{miles2007multidetector}. According to the convolutional model for cerebral perfusion parametric mapping \cite{miles2007multidetector,RN125,RN124,RN123}, the relationship between the time-resolved image matrix $\mathbf{X}$ and the flow-scaled residual matrix $\mathbf{C}$ can be represented by the following linear model:
	\begin{align}
		\label{eq:convolution-model}
		\mathbf{X} &= \Delta t \cdot \mathbf{C}\mathbf{B}^{\text{T}} \\
		B_{i,j} &=
		\begin{cases}
			h_{\text{AIF}}(t_{i-j+1}), & j \leq i \\[5pt]
			0, & j > i
		\end{cases} \\
		C_{i,j} &=
		\begin{cases}
			\rho \cdot \text{CBF}_i, & t_j < \mathcal{T}_{0_{i}}\\[2pt]
			\rho \cdot \text{CBF}_i \cdot \exp\left( - \frac{\left( t_j - \mathcal{T}_{0_{i}} \right)}{h} \right), & t_j \geq \mathcal{T}_{0_{i}} 
		\end{cases}
	\end{align}
	Where $\mathbf{B}\in\mathbb{R}^{T\times T}$, %a block-circulant matrix that reduces the influence of the bolus delay in the deconvolution process \cite{wittsack2008ct}, 
	$B_{i,j}$ denotes the matrix element of $\mathbf{B}$, $h_{\text{AIF}}(t_i)$ denotes the digitized arterial input function \cite{RN46}. $\Delta t$ denotes the unit time interval, and $h$ denotes a unit conversion constant with a unit of seconds. $C_{i,j}$ denotes the matrix element of $\mathbf{C}\in\mathbb{R}^{N\times T}$ which represents the flow-scaled residual matrix stacking up each flow-scaled residual function values of all image voxels associated with the $j$-th time frame, $\mathbf{c}_j\in\mathbb{R}^{N\times 1}$, along its column direction. The matrix form $\mathbf{C}\mathbf{B}^{\text{T}}$ represents the digitized convolutional process of the arterial input function $h_{\text{AIF}}(t)$ and the flow-scaled residual function for each image voxel independently. $\rho$ denotes the nominal density of cerebral tissues. $\text{CBF}_i$ and $\mathcal{T}_{0_{i}}$ denote the $i$-th element of the two perfusion parametric images to be reconstructed, $\mathbf{\text{CBF}}$ and $\mathbf{\mathcal{T}}_{0}$, respectively.
	
	According to the convolutional model for cerebral perfusion parametric mapping, the perfusion parametric images $\mathbf{\text{CBV}}$ and $\mathbf{\text{MTT}}$ can be derived using the following equations once the optimal CBF$^*$ and $\mathbf{\mathcal{T}}_{0}^*$ are obtained:
	\begin{align}
		\text{CBV}^*_i &\approx \frac{1}{\rho} \cdot \sum_{j=1}^{T} [\mathbf{C}(\text{CBF}^*, \mathbf{\mathcal{T}}_{0}^*)]_{i,j} \cdot \Delta t \\[5pt]
		\text{MTT}^*_i &= \frac{\text{CBV}^*_i}{\text{CBF}^*_i}
	\end{align}
	$\mathbf{\text{CBV}}^*_{i}$ and $\mathbf{\text{MTT}}^*_{i}$ denote the $i$-th element of the two obtained perfusion parametric images, $\mathbf{\text{CBV}}$ and $\mathbf{\text{MTT}}$, respectively.
	
	\subsection{Quantitative Cerebral Perfusion Image Reconstruction Directly From Measured Sinogram Data}
	
	The two perfusion parametric images to be reconstructed, $\mathbf{\text{CBF}}$ and $\mathbf{\mathcal{T}}_{0}$, can be obtained by solving the following optimization problem:
	\begin{align}
		\label{eq:least-square}
		\mathbf{U}^* &= \argmin_{\mathbf{U}} \frac{1}{2} ||\vec{\mathbf{Y}}-\overrightarrow{\bar{\mathbf{Y}}(\mathbf{U})}||_\mathbf{W}^2, \\ \nonumber
		\mathbf{U} &:= \left[ \mathbf{\text{CBF}}, \mathbf{\mathcal{T}}_{0}\right], 
		%&:= \argmin_\mathbf{C} L(\mathbf{C})
	\end{align}
	where $\mathbf{W}\in\mathbb{R}^{MT_0\times MT_0}$ denotes the statistical weighting matrix and $\vec{\mathbf{Y}}\in\mathbb{R}^{MT_0\times 1}$ denotes the measured data stacking the measurements associated with each scan along its column direction. 
	
	\subsection{Subject-Specific Conditional Generative Models for Perfusion Parametric Image Representation}
	
	To avoid potentially biased regularization design for perfusion parametric estimation, the perfusion parametric images in the proposed TRAINER have been represented as a subject-specific conditional generative model trained under the constraint of the subject's own measured sinogram data. We parameterize the unknowns ($\mathbf{\text{CBF}}$ and $\mathbf{\mathcal{T}}_{0}$) as follows:
	\begin{align}
		\label{eq:generative-model}
		\mathbf{U} &= \mathcal{G}_\mathbf{\Theta} (\mathbf{Z}),
	\end{align}  
	where $\mathcal{G}_\mathbf{\Theta}$ denotes the parameterized perfusion images, $\mathbf{\text{CBF}}$ and $\mathbf{\mathcal{T}}_{0}$, from the noise image sampled from the uniform distribution, $\mathbf{Z}\in\mathbb{R}^{N\times 2}$. Inspired by the deep image prior \cite{ulyanov2018deep}, we represent $\mathcal{G}_\mathbf{\Theta}$ by typical convolutional neural networks, whose model parameters are denoted by $\mathbf{\Theta}$. 
	
	\subsection{The Joint Optimization Problem in TRAINER}
	
	With the designed conditional generative model, we propose to estimate the perfusion parametric images $\mathbf{U}$ and the associated model parameters $\mathbf{\Theta}$ by fitting the model to the measured sinogram data by incorporating the forward model for time-resolved CT data acquisition and convolutional model for cerebral perfusion parametric estimation by solving the following constrained optimization problem: 
	\begin{align}
		\label{eq:joint-optimization}
		\mathbf{U}^* &= \argmin_{\mathbf{U}} \frac{1}{2} ||\vec{\mathbf{Y}}-\overrightarrow{\bar{\mathbf{Y}}(\mathbf{U})}||_\mathbf{W}^2, \\
		s.t. ~~~ \mathbf{U}  &= \mathcal{G}_\mathbf{\Theta} (\mathbf{Z}).
	\end{align}
	
	We use the augmented Lagrangian formula to convert the constrained problem into a unconstrained problem:
	\begin{align}
		\label{eq:unconstrained-optimization}
		\mathcal{L}_{\lambda}(\mathbf{U},\mathbf{\Theta},\mathbf{V})&=\frac{1}{2} ||\vec{\mathbf{Y}}-\overrightarrow{\bar{\mathbf{Y}}(\mathbf{U})}||_\mathbf{W}^2 \\ \nonumber
		&+\frac{\lambda }{2}||\mathbf{U}-\mathcal{G}_\mathbf{\Theta} (\mathbf{Z})+\mathbf{V}||^2,
	\end{align}
	Where $\mathbf{V}$ denotes the dual variable of $\mathbf{U}$, and $\lambda$ controls the trade-off between data fitting term and the regularization term. 
	
	The unconstrained problem can be efficiently solved by the alternating direction method of multipliers (ADMM) algorithm iteratively through the following three sub-problems:
	\begin{itemize}
		\item Perfusion Parametric Image Reconstruction Sub-problem:
		\begin{align}
			\label{eq:subproblem-1}
			\mathbf{U}^{(k+1)} &=\argmin_{\mathbf{U}} \mathcal{L}_{\lambda}(\mathbf{U},\mathbf{\Theta}^{(k)},\mathbf{V}^{(k)}), \\
			&=\argmin_{\mathbf{U}} \frac{1}{2} ||\vec{\mathbf{Y}}-\overrightarrow{\bar{\mathbf{Y}}(\mathbf{U})}||_\mathbf{W}^2 \\ \nonumber
			&+\frac{\lambda }{2}||\mathbf{U}-\mathcal{G}_{\mathbf{\Theta}^{(k)}} (\mathbf{Z})+\mathbf{V}^{(k)}||^2
		\end{align}
		
		\item Conditional Generative Model Training Sub-problem:
		\begin{align}
			\label{eq:subproblem-2}
			\mathbf{\Theta}^{(k+1)} &= \argmin_{\mathbf{\Theta}}\mathcal{L}_{\lambda}(\mathbf{U}^{(k+1)},\mathbf{\Theta},\mathbf{V}^{(k)}), \\
			&=\argmin_{\mathbf{\Theta}} ||\mathbf{U}^{(k+1)}+\mathbf{V}^{(k)}-\mathcal{G}_\mathbf{\Theta} (\mathbf{Z})||^2
		\end{align}
		
		\item Dual Variable Updating Sub-problem:
		\begin{align}
			\mathbf{V}^{(k+1)} &= \mathbf{V}^{(k)} + \mathbf{U}^{(k+1)} - \mathcal{G}_{\mathbf{\Theta}^{(k+1)}}(\mathbf{Z})
			\label{eq:ADMM3}
		\end{align}
	\end{itemize}
	
	\subsection{Numerical Optimization Algorithm}

	\subsubsection{Solving sub-problem Eq.~(\ref{eq:subproblem-1})}
	
	The optimization of sub-problem Eq.~(\ref{eq:subproblem-1}) can be technically challenging. As we can observe, the cost function in Eq.~(\ref{eq:subproblem-1}) is highly nonlinear and non-convex with respect to the unknown perfusion parametric images. Here, we choose to use the Adam optimizer \cite{kingma2014adam} to minimize the cost function defined in Eq.~(\ref{eq:subproblem-1}). The Adam optimizer is particularly suitable for this task due to its ability to handle non-smooth objective functions and dynamically adapt learning rates to facilitate faster convergence.
	
	\subsubsection{Solving sub-problem Eq.~(\ref{eq:subproblem-2})}
	
	Sub-problem Eq.~(\ref{eq:subproblem-2}) is a nonlinear least square problem and it has the same form as the conventional supervised training problem, using $\mathbf{U}^{(k+1)}+\mathbf{V}^{(k)}$ as the training labels. Gradient descent has been employed to solve this supervised training problem. 
	
	%Following Eq.~(\ref{eq:subproblem-2}), we regularize the problem by considering the \edit{parametrization} $\mathcal{G}_{\mathbf{\Theta}}$ and optimizing the resulting energy w.r.t. $\mathbf{\Theta}$. Optimization still applies gradient descent as the optimizer, exploiting the fact that both the neural network and the most common downsampling operators, such as Lanczos, are differentiable \cite{dmitry2020deep}.
	%, leveraging the differentiability properties of both the neural network and common downsampling operators such as Lanczos.
	
	\subsubsection{Pseudo-code to Implement TRAINER}
	
	To implement TRAINER, we first need to initialize $\mathbf{U}$, $\mathbf{V}$, and $\mathbf{\Theta}$, and then optimize these arguments iteratively. 
	\begin{algorithm}[H]
		\caption{TRAINER}
		\label{alg:TRAINER}
		\begin{algorithmic}[1] 
			\STATE \textbf{Input:} Line integral data $\vec{\mathbf{Y}}$, Initial $\mathbf{CBF}$ and $\mathbf{\mathcal{T}}_{0}$
			\STATE \textbf{Output:} $\mathbf{U} = \{\mathbf{CBF}, \mathbf{\mathcal{T}}_{0}\}$
			\STATE \textbf{Initialize:} $\mathbf{V}^{(0)}, \mathbf{\Theta}^{(0)}, \mathbf{Z}, \lambda, \eta, \alpha, \varepsilon$
			\WHILE{$k \leq \text{Num Of ADMM}$ and $\mathcal{L}_{\lambda}(\mathbf{U}^{(k)}, \mathbf{\Theta}^{(k)}, \mathbf{V}^{(k)}) \geq \varepsilon$}  \label{alg:outer_loop}
			\STATE $\mathbf{U}^{(0)} \gets \mathbf{U}^{(k)}$ 
			\STATE Update $\mathbf{U}^{(k+1)}$ using Adam optimizer:
			
			\FOR{$q = 0$ to $Q$}
			\STATE Compute $\mathbf{g}_{\mathbf{U}}^{(q)} = \nabla_{\mathbf{U}} \bigg[ \frac{1}{2} \|\vec{\mathbf{Y}} - \overrightarrow{\bar{\mathbf{Y}}(\mathbf{U})}\|_\mathbf{W}^2 + \frac{\lambda}{2} \|\mathbf{U} - \mathcal{G}_{\mathbf{\Theta}^{(k)}}(\mathbf{Z}) + \mathbf{V}^{(k)}\|^2 \bigg]$
			\STATE $\mathbf{U}^{(q+1)} = \text{AdamUpdate}(\mathbf{U}^{(q)}, \mathbf{g}_{\mathbf{U}}^{(q)}, \eta)$
			\ENDFOR
			\vspace{0.2\baselineskip}
			\STATE $\mathbf{U}^{(k+1)} \gets \mathbf{U}^{(Q)}$
			\STATE Update $\mathbf{\Theta}^{(k+1)}$ via gradient descent:
			\FOR{$p = 0$ to $P$}
			\STATE Compute $\mathbf{g}_{\mathbf{\Theta}}^{(p)} = \nabla_{\mathbf{\Theta}} \|\mathbf{U}^{(k+1)} + \mathbf{V}^{(k)} - \mathcal{G}_\mathbf{\Theta}(\mathbf{Z})\|^2$
			\vspace{0.2\baselineskip}
			\STATE $\mathbf{\Theta}^{(p+1)} = \mathbf{\Theta}^{(p)} - \alpha \mathbf{g}_{\mathbf{\Theta}}^{(p)}$
			\ENDFOR
			\STATE $\mathbf{\Theta}^{(k+1)} \gets \mathbf{\Theta}^{(P)}$
			\STATE Update $\mathbf{V}^{(k+1)}$:
			\vspace{0.2\baselineskip}
			\STATE $\mathbf{V}^{(k+1)} = \mathbf{V}^{(k)} + \mathbf{U}^{(k+1)} -\mathcal{G}_{\mathbf{\Theta}^{(k+1)}}(\mathbf{Z})$
			\STATE $k \gets k + 1$
			\ENDWHILE
			\STATE Finalize and output results
		\end{algorithmic}
	\end{algorithm}
	Notably, the iteration counts for $\text{Num Of ADMM}$, $P$, and $Q$, which are hyperparameters, are empirically set to $\text{Num Of ADMM}=700$, $P=50$, and $Q=2000$. These values can be modified to accommodate specific imaging tasks, data acquisition and contrast agent injection protocols, or other data-related conditions.
	%The perfusion parametric reconstruction sub-problem (Eq.~(\ref{eq:subproblem-1})) has been solved by a iterative process typically performed for 300 iterations to obtain $\mathbf{U}^{(k+1)}$. The conditional generative model training sub-problem (Eq.~(\ref{eq:subproblem-2})) has been solved by a iterative process typically performed for 2000 iterations to obtain $\mathbf{\Theta}^{(k+1)}$. Finally, the Lagrangian multiplier $\mathbf{V}$ is updated, and this iterative process continues until convergence criteria are met. 

	%%%%%%%%%%%%%%%%%%%%%%%%%%%%%%%%%%%%%%%%%%%%%%%%%%%%%%%%%%%%%%%%%%%%%%%%%%%%%%%%%%%%%%%%%%%%%%%%%%%%%%%%%%%

	\subsubsection{Network Architecture}
	
\begin{figure*}[t] 
		\centering  
		\includegraphics[width=1\linewidth]{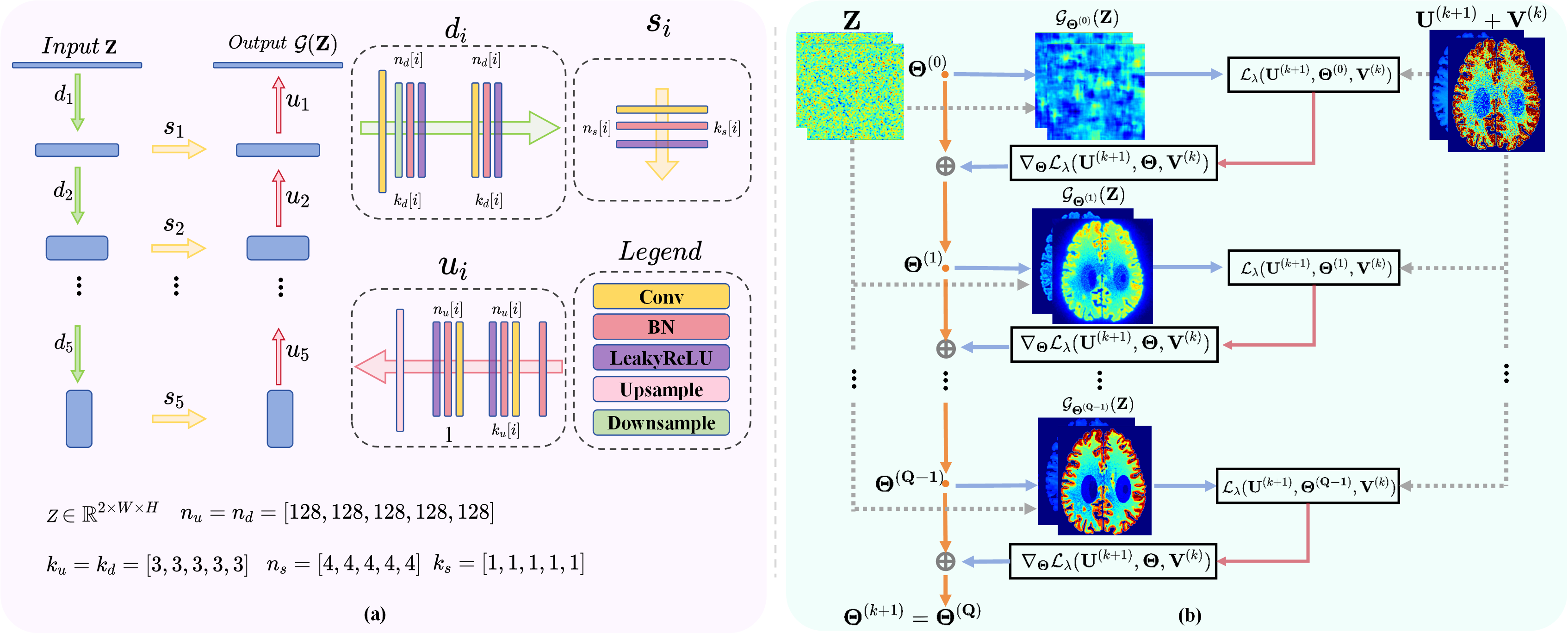} 
		\caption{(a) The architecture of $\mathcal{G}$. $n_u[i]$, $n_d[i]$, and $n_s[i]$ denote the number of filters for upsampling, downsampling, and skip connections at the $i$-th depth respectively. Parameters $k_u[i]$, $k_d[i]$, and $k_s[i]$ denote the kernel sizes at the $i$-th depth respectively. (b) Conditional image generation. During each iteration, the network parameters $\mathbf{\Theta}$ determine the image $\mathcal{G}_\mathbf{\Theta}(\mathbf{Z})$, and the mapping $\mathcal{G}$ is represented as a deep convolutional neural network parameterized by $\mathbf{\Theta}$.}  
		\label{fig:network}  
	\end{figure*}

	The architecture of $\mathcal{G}_\mathbf{\Theta}$ is illustrated in Fig.~\ref{fig:network} (a). The architecture is designed based on an encoder-decoder structure with skip connections. Parameters $n_u[i]$, $n_d[i]$, and $n_s[i]$ denote the number of filters for upsampling, downsampling, and skip connections at the $i$-th depth, respectively. Parameters $k_u[i]$, $k_d[i]$, and $k_s[i]$ denote the kernel sizes at the $i$-th depth, respectively \cite{dmitry2020deep}. Bilinear interpolation is employed for the upsampling process. We use LeakyReLU \cite{he2015delving} to introduce the non-linearity. During the iterative training process of network parameters $\mathbf{\Theta}$, we add additive normal noise with a mean of zero and a standard deviation of $\sigma_p = \frac{1}{30}$ to input $\mathbf{Z}$ for perturbation. Experimental results show that this method yields superior performance. Additionally, previous research indicates that the optimization process tends to become unstable as the loss decreases and approaches a specific threshold. This instability is characterized by a notable increase in the loss value and the appearance of blurriness in the generated image $\mathcal{G}_\mathbf{\Theta}$. To make the training process more stable, we revert the parameters to those from the previous iteration if the loss difference between two consecutive iterations exceeds a specified threshold \cite{dmitry2020deep}.
	
	\subsection{Implementation Details}
	
	%	\subsubsection{Obtaining AIF}
	%	
	%	To apply TRAINER, it is necessary to accurately reconstruct the AIF and then reconstruct the perfusion parametric images using the AIF obtained. In both the numerical simulation studies and in vivo animal subject studies, we deliberately applied the same AIF across all baseline methods and TRAINER to establish fair comparisons. By decoupling the AIF reconstruction from the perfusion parametric image reconstruction, the contributions of different parametric image reconstruction approaches can be highlighted. In numerical simulation studies, we applied the ground truth AIF to generate the corresponding parametric images using all baseline methods and TRAINER. In in vivo animal subject studies, since the ground truth AIF is unavailable, we used our previously developed eSMART-RECON \cite{li2019enhanced} to generate the AIF and then used to generate the corresponding parametric images using all baseline methods and TRAINER.
	
	\subsubsection{Initialization and Hyperparameter Setting}
	
	We first initialized the iterative sequence via the following empirical schemes to obtain the warm initialization of $\mathbf{U}$. The value of CBF for each pixel is initialized as 40 mL/100g/min and $\mathbf{\mathcal{T}}_{0}$ for each pixel is initialized as 5 seconds because such initialization helps the iterative process converge more quickly to the expected solution. $\mathbf{\Theta}$ is initialized as random values uniformly distributed between 0 and 1, while $\mathbf{V}$ is initialized as 0. The variable $\mathbf{Z}$ is initialized within the interval $[0, 1/10)$. To optimize the performance of TRAINER, the hyperparameters $\lambda$, $\alpha$, $\eta$ for numerical simulation studies are empirically set to 0.01, 0.01, and 0.001, respectively. %For in vivo animal subject studies, the hyperparameters $\lambda$, $\alpha, \eta$ are empirically set to 0.012, 0.008, and 0.001, respectively. Note that, these hyperparameters need to be optimized according to the imaging task, data acquisition, and contrast injection protocols and/or other data conditions etc.
	
	%Hyperparameters $\lambda$, $\alpha, \eta$ are empirically set to 0.0015, 0.01, and 0.001 respectively, for optimal performance of TRAINER.
	
	\subsubsection{Other Remarks}
	
	TRAINER is implemented using Pytorch 2.1.0 on a Linux system equipped with four NVIDIA GeForce RTX 3090.
	
	\section{Materials and Methods to Validate and Evaluate TRAINER}
	
	\subsection{Numerical Simulation Studies with Known Ground Truth}
	\label{sec:materials-phantom}
	
	The design of the numerical phantom in this study was based on a digital anthropomorphic perfusion phantom \cite{RN183}. Regions corresponding to artery, healthy tissue, cerebrospinal fluid, penumbra and ischemic core are defined in the phantom design software toolkit. The residual functions are defined according to programmed cerebral blood flow and mean transit time values. The ground truth TAC of each pixel is then calculated according to the indicator dilution theory. The data acquisition geometry was designed to simulate the scans performed in the animal subject study. 
	
	To be specific, the angular span of each scan is $360^{\circ}$. The data acquisition time per scan was $8$ seconds, and a total of $200$ projections were acquired for each scan. Each projection includes $380\times 1$ measured line integrals. An image matrix size of $256\times 256$ was used to reconstruct each time frame. The reconstructed image has the isotropic spatial resolution of $1\text{ mm}\times 1\text{ mm}$. During the total data acquisition duration, about $64$ seconds, two mask scans (without contrast injection) plus a total of eight bidirectional rotational scans with contrast injection (four forward rotations and four reverse rotations) were simulated.
	
	Ideal projection data were generated using a ray-driven implementation of fan-beam forward projection. Projections with quantum noise were then produced by sampling from a Poisson distribution for each projection ray of the ideal data. For CBCT perfusion data acquisitions, an entrance photon number of \( I_0 = 1 \times 10^6 \) photons per ray was used to simulate the full exposure level. Entrance photon number per ray $I_{0}=1\times 10^{5},5\times 10^{4}, 1\times 10^{4}$ were used to simulate lower exposure levels respectively. 
	
	%\edit{For CBCT perfusion data acquisitions, an entrance photon number of \( I_0 = 1 \times 10^6 \) photons per ray was used, which is deliberately set significantly higher than the standard dose. This higher dose is necessary to reduce quantum noise and improve the signal-to-noise ratio (SNR) during data acquisition, which is crucial for accurate perfusion imaging.} 
	%For CBCT perfusion data acquisitions, entrance photon numbers $I_0=1\times 10^6$ photons per ray were used, such that, noise levels were deliberately selected to match the image noise measured for standard exposure level. 
	%We denote this exposure level as the full exposure level. Entrance photon number per ray $I_{0}=1\times 10^{5},5\times 10^{4}, 1\times 10^{4}$ were used to simulated lower exposure levels respectively. 

	\subsection{Baseline Methods for Performance Comparison} 
	
	Four baseline methods are compared: %standard truncated singular value decomposition (sSVD) \cite{ostergaard1996high}, block-circulant truncated SVD (bSVD) \cite{wu2003tracer}, Tikhonov regularization% 
	the Singular Value Decomposition (SVD)-based with Tikhonov regularization algorithms \cite{fieselmann2011deconvolution}, and Tensor Total-Variation Regularization (TTV) \cite{fang2015robust}, temporal recovery technique (TRT) \cite{tang2013novel}, and eSMART-RECON \cite{li2019enhanced}. For the (SVD)-based with Tikhonov regularization algorithms, threshold value of 0.15 (15\% of the maximum singular value) is empirically chosen to achieve optimal performance. These methods are widely recognized as regularized deconvolution techniques for CT perfusion imaging and are extensively incorporated into commercial medical software \cite{kudo2010differences}. TTV \cite{fang2015robust} applies total variation regularization to the impulse response function, integrating spatial vascular structure correlation and temporal blood signal continuity. This algorithm converges rapidly and exhibits low computational complexity. In the implementation of the TTV algorithm in this study, both the regularization parameters for the temporal and spatial dimensions, denoted as $\gamma$, were set to 0.001, with a total of 60 iterations performed. TRT algorithm introduces a novel temporal recovery technique designed to restore the time attenuation curves in C-arm CBCT perfusion imaging, addressing the limitations of existing C-arm CBCT systems in terms of data acquisition speed and temporal sampling density. In the implementation of the TRT algorithm in this study, the correction factor was set to 0.5. eSMART improves the temporal resolution of C-arm CBCT via incorporating the prior knowledge of periodicity of the limited-view artifacts in the C-arm CBCT perfusion data acquisition scheme. For each time point, the eSMART algorithm selects the 10 adjacent frames of projection data as input, and performs 15 iterations. A relaxation factor of 0.06 is introduced during the iteration process to optimize both iteration efficiency and the quality of the reconstructed images. To improve image quality and quantitative accuracy at lower exposure levels, noise reduction in the time-resolved images reconstructed by each baseline method was achieved through spatial and temporal domain filtering. Specifically, spatial filtering was performed using a Gaussian filter with a standard deviation of 1 and a kernel size of \(5 \times 5\), while temporal filtering was conducted with a Gaussian filter having a standard deviation of 1 and a kernel size of \(3 \times 1\).  Perfusion parametric images of each baseline method were then generated using the corresponding denoised time-resolved images.
	
	%\edit{The specific implementation details of the eSMART algorithm in this study are as follows: The eSMART algorithm first employs low temporal-resolution C-arm projection data and applies a block-based backprojection algorithm for interpolation, generating preliminary reconstruction images at multiple time points. Simultaneously, the method integrates the Simultaneous Algebraic Reconstruction Technique (SART) for iterative reconstruction with limited-angle projection data. For each time point, the algorithm selects the 10 adjacent frames of projection data as input, performs 15 iterations, and subsequently applies Singular Value Decomposition (SVD) to mitigate artifacts arising from limited-angle sampling. A relaxation factor of 0.06 is introduced during the iteration process to optimize both iteration efficiency and the quality of the reconstructed images.} 
	
	\subsection{Metrics for Quantitative Performance Assessment} 
	
	The study employs relative Root Mean Square Error (rRMSE) and Peak Signal-to-Noise Ratio (PSNR) as evaluation metrics.
	\begin{equation}
		\mathrm{rRMSE}=\frac{\sqrt{\frac{1}{n} \sum_{i=1}^{n}\left(y_{i}-y_{i}^{\text{truth}}\right)^{2}}}{\max y^{\text{truth}}}  \times 100 \%,
	\end{equation}
	where $y$ denotes the image being assessed, $y^{\text{truth}}$ denotes the ground truth image, $\max y^{\text{truth}}$ denotes the maximum pixel value of the ground truth image, and the subscript $i$ denotes the $i$-th voxel index of the image.
	
	\begin{equation}
		\text{PSNR} = 10 \cdot \log_{10} \left( \frac{{\text{MAX}^2}}{{\text{MSE}}} \right), 
	\end{equation}
	where \(\text{MAX}\) denotes the maximum possible pixel value of the image , and \(\text{MSE}\) denotes the Mean Squared Error between the ground truth image \(y^{\text{truth}}\) and the estimated image \({y}\), which is calculated as:
	\begin{equation}
		\label{eq:mse}
		\text{MSE} = \frac{1}{n} \sum_{i=1}^{n} (y^{\text{truth}}_{i} - {y}_{i})^2.
	\end{equation}
	
	%%%%%%%%%%%%%%%%%%%%%%%%%%%%%%%%%%%%%%%%%%%%%%%%%%%%%%%%%%%%%%%%%%%%%%%%%%%%%%%%%%%%%%%%%%%%%%%%%%%%%
	
	\section{Results}
	
	\subsection{Validation of Numerical Optimization}
	
\begin{figure}[htbp]
		\centering
		\includegraphics[width=1\linewidth]
		{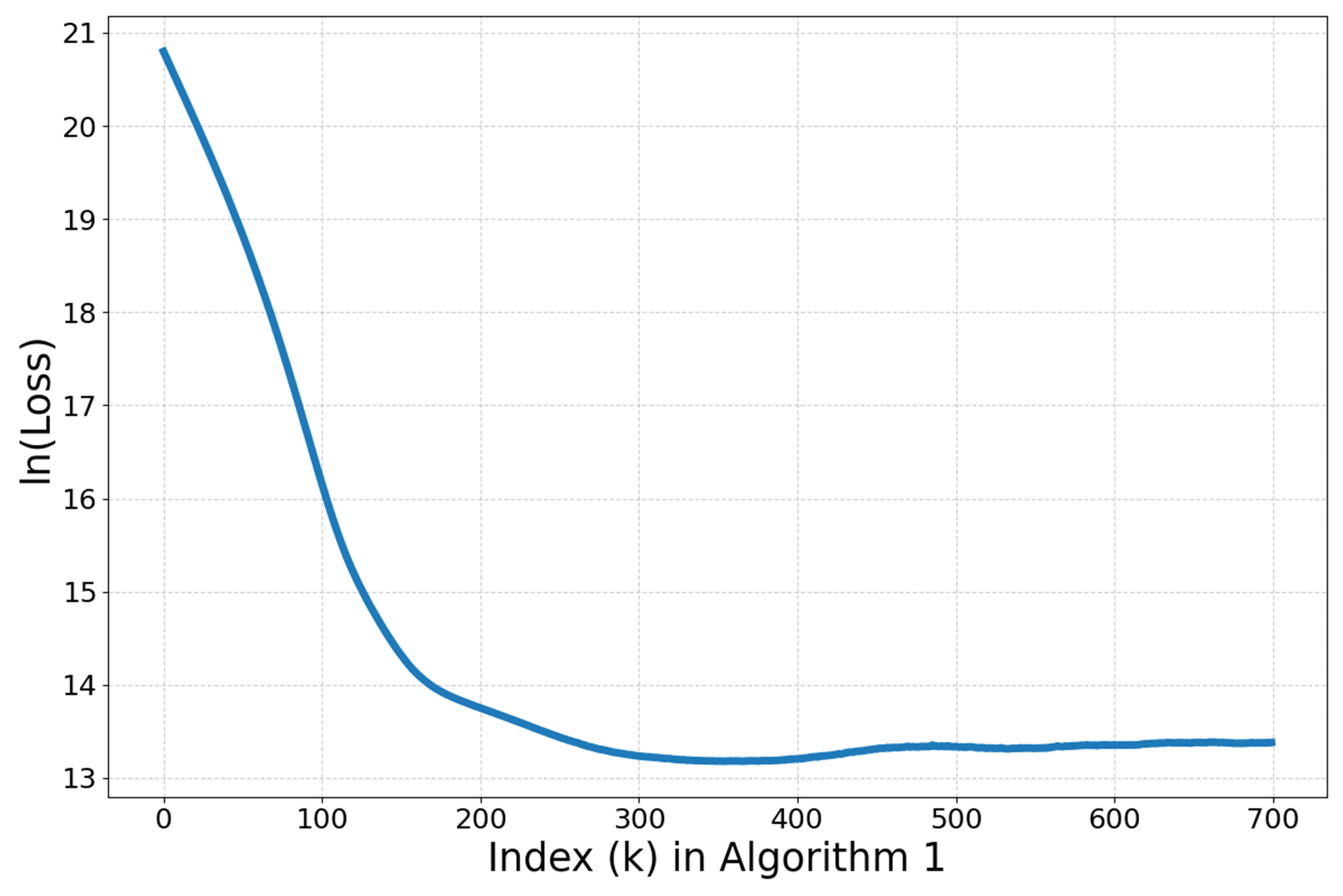}
		\caption{The change of loss function value (Eq.~\ref{eq:unconstrained-optimization}) with respect to each iteration.}
		%\caption{The change of loss function value and error value of predicted perfusion parametric images with respect to each epoch at temporal resolution of $dt=8$ sec under different radiation exposure level conditions.}
		\label{fig:loss}
	\end{figure}

	Empirical convergence of the optimization process of TRAINER is shown as the change of loss function value (Eq.~\ref{eq:unconstrained-optimization}) with respect to each iteration in Fig.~\ref{fig:loss}. As shown in Fig.~\ref{fig:loss}, the loss function value quasi-monotonically decreases during the optimization. The plateau of the total loss values indicates the empirical convergence of the numerical optimization process.
	
	\subsection{Qualitative Assessment of Perfusion Parametric Images}
	
\begin{figure}[htbp]
		\centering
		\includegraphics[width=1\linewidth]
		{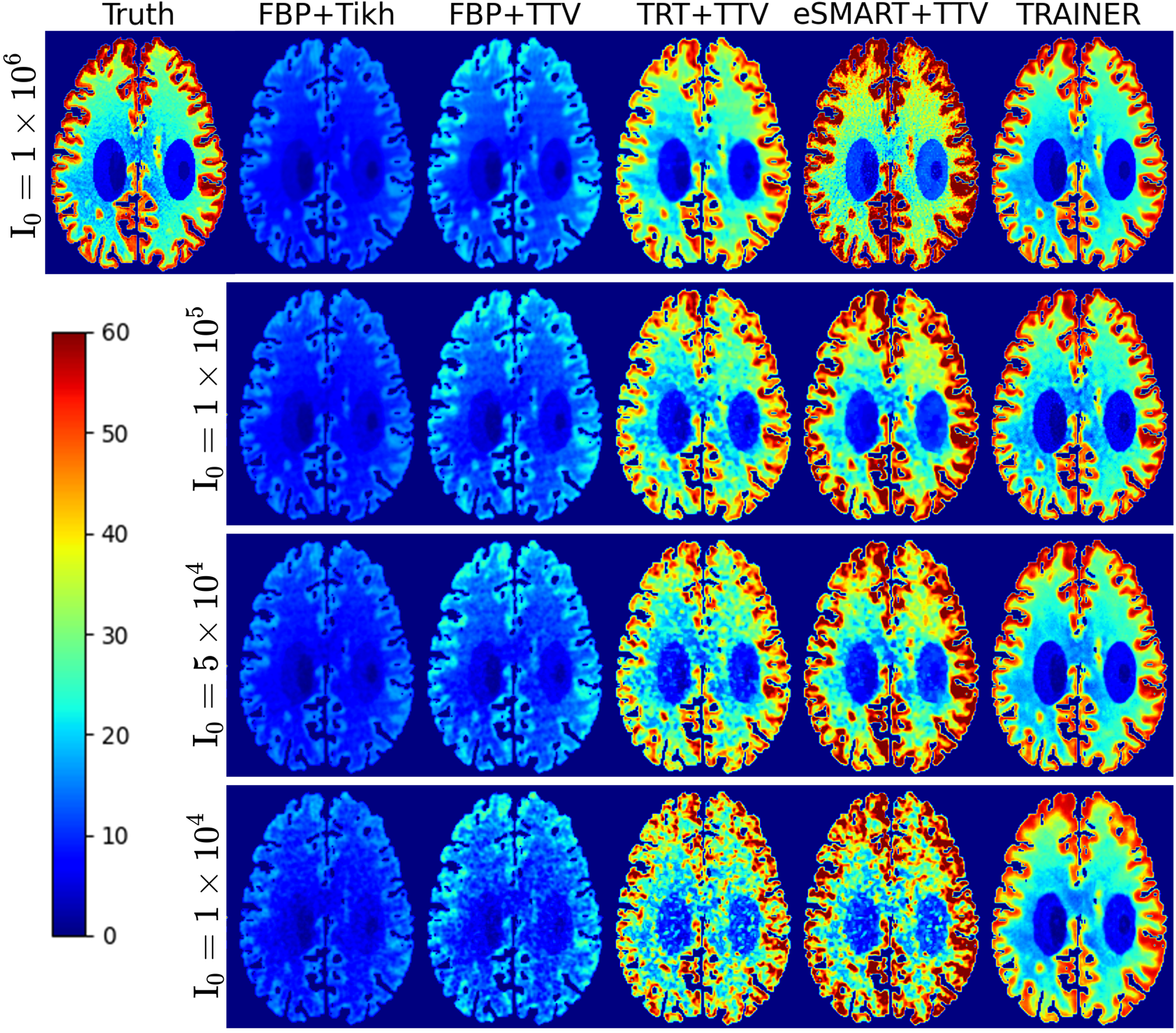}
		\caption{CBF images of numerical phantom generated from the ground truth, baseline methods, and TRAINER. Data were simulated at gantry rotation speed of $dt=8$ sec and four exposure levels. Display windows are shown as the color bar. All images are shown with a W/L: 60/30.}
		\label{fig:cbf}
	\end{figure}

\begin{figure}[htbp]
		\centering
		\includegraphics[width=1\linewidth]
		{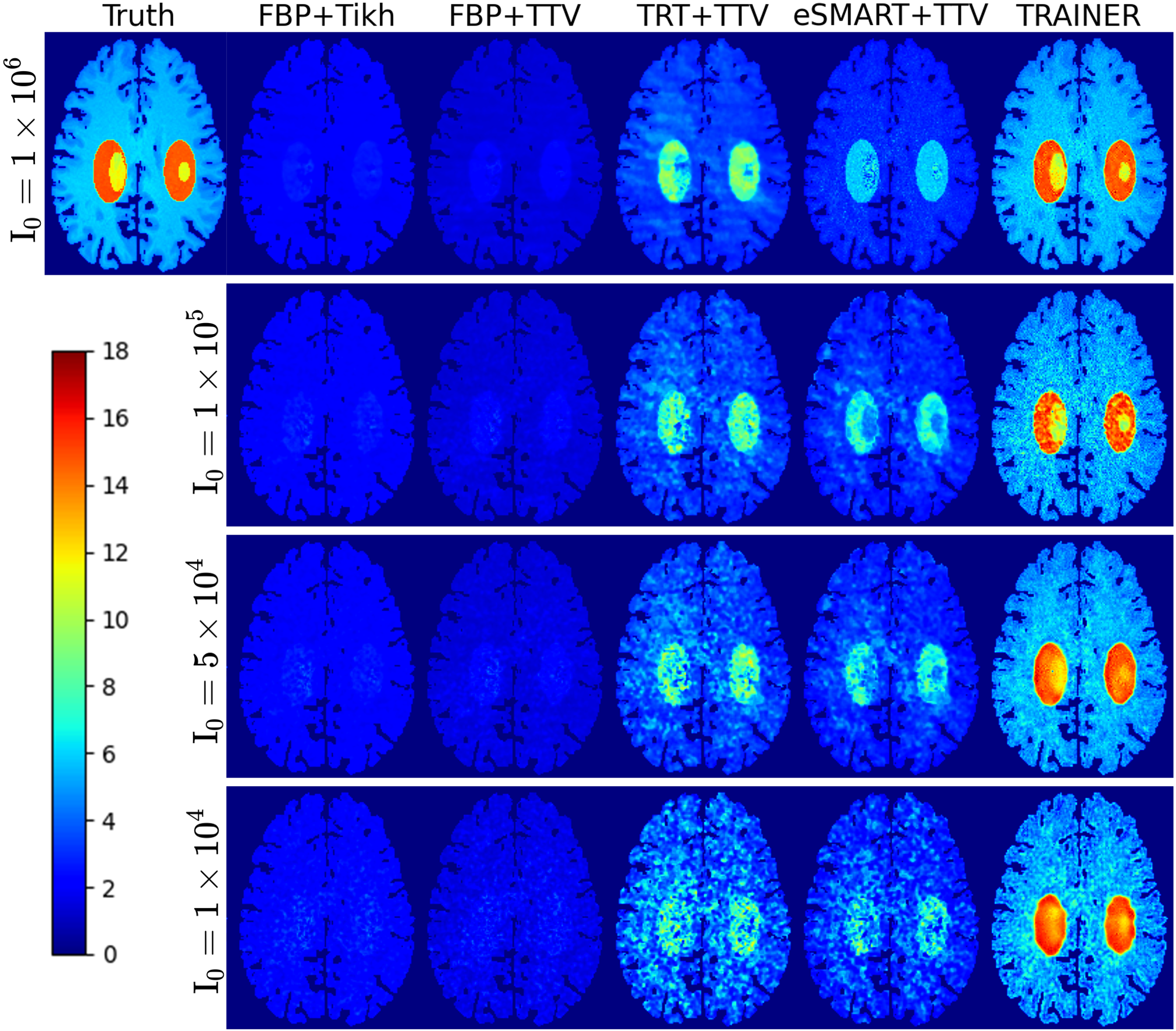}
		\caption{MTT images of numerical phantom generated from the ground truth, baseline methods, and TRAINER. Data were simulated at gantry rotation speed of $dt=8$ sec and four exposure levels. Display windows are shown as the color bar. All images are shown with a W/L: 18/9.}
		\label{fig:mtt}
	\end{figure}

	Quantitative perfusion images reconstructed at the gantry rotation speed of $dt=8$ sec and four exposure levels are shown in Fig.~\ref{fig:cbf} and Fig.~\ref{fig:mtt}. For CBF images, at the full exposure level, compared with the ground truth, FBP+Tikh and FBP+TTV underestimate healthy tissue while overestimate diseased tissues (penumbra and ischemic core). TRT+TTV underestimates but eSMART+TTV overestimates healthy tissue. In addition, both TRT+TTV and eSMART+TTV overestimate diseased tissues. In contrast, TRAINER can accurately reconstruct CBF values for all types of tissues. When the exposure level is getting lower, penumbra and ischemic core are not distinguished in baseline methods. Error and noise level of CBF images reconstructed by baseline methods are getting larger. For MTT images, at the full exposure level, compared with the ground truth, baseline methods underestimate all tissue types. Penumbra and ischemic core are not distinguished in baseline methods even at the full exposure level. TRAINER accurately estimate the value of all types of tissues. When the exposure level is getting lower, error and noise level of MTT images reconstructed by baseline methods are getting larger. Results shown in Fig.~\ref{fig:cbf} and Fig.~\ref{fig:mtt} demonstrated that TRAINER can accurately reconstruct quantitative perfusion images at gantry rotation speed of $dt=8$ sec even at the lowest exposure level for all tissue types while other methods cannot.
	
	%For better visualizing quality of penumbra and ischemic core, zoom-in CBF images are shown in Fig.~\ref{fig:cbf-roi}. 
	%For better visualizing quality of penumbra and ischemic core, zoom-in MTT images are shown in Fig.~\ref{fig:mtt-roi}. 
	
	%\begin{figure}[htbp]
	%            \centering
	%            \includegraphics[width=1\linewidth]
	%            {figures/CBF_ROI.png}
	%            \caption{Zoom-in CBF images of numerical phantom generated from ground truth, baseline methods, and TRAINER. Data were simulated at gantry rotation speed of $dt=8$ sec and four exposure levels. Display windows are shown as the color bar. All images are shown with a W/L: 60/30. The rRMSE for each of these images was calculated.}
	%            \label{fig:cbf-roi}
	%\end{figure}
	
	%\begin{figure}[htbp]
	%	\centering
	%	\includegraphics[width=1\linewidth]
	%	{figures/MTT_ROI.png}
	%	\caption{Zoom-in MTT images of numerical phantom generated from ground truth, baseline methods, and TRAINER. Data were simulated at gantry rotation speed of $dt=8$ sec and four exposure levels. Display windows are shown as the color bar. All images are shown with a W/L: 18/9. The rRMSE for each of these images was calculated.}
	%	\label{fig:mtt-roi}
	%\end{figure}
	
	To validate the effectiveness of the proposed TRAINER and also explain why other baseline methods cannot obtain accurate quantitative perfusion images, we presented the time attenuation curves (TACs) of the reconstructed time-resolved images using TRAINER and compared them to those obtained from the baseline methods.%, which employ Filtered Back Projection (FBP) reconstruction plus deconvolution to estimate parametric images
	
\begin{figure}[htbp]
		\centering
		\includegraphics[width=1\linewidth]
		{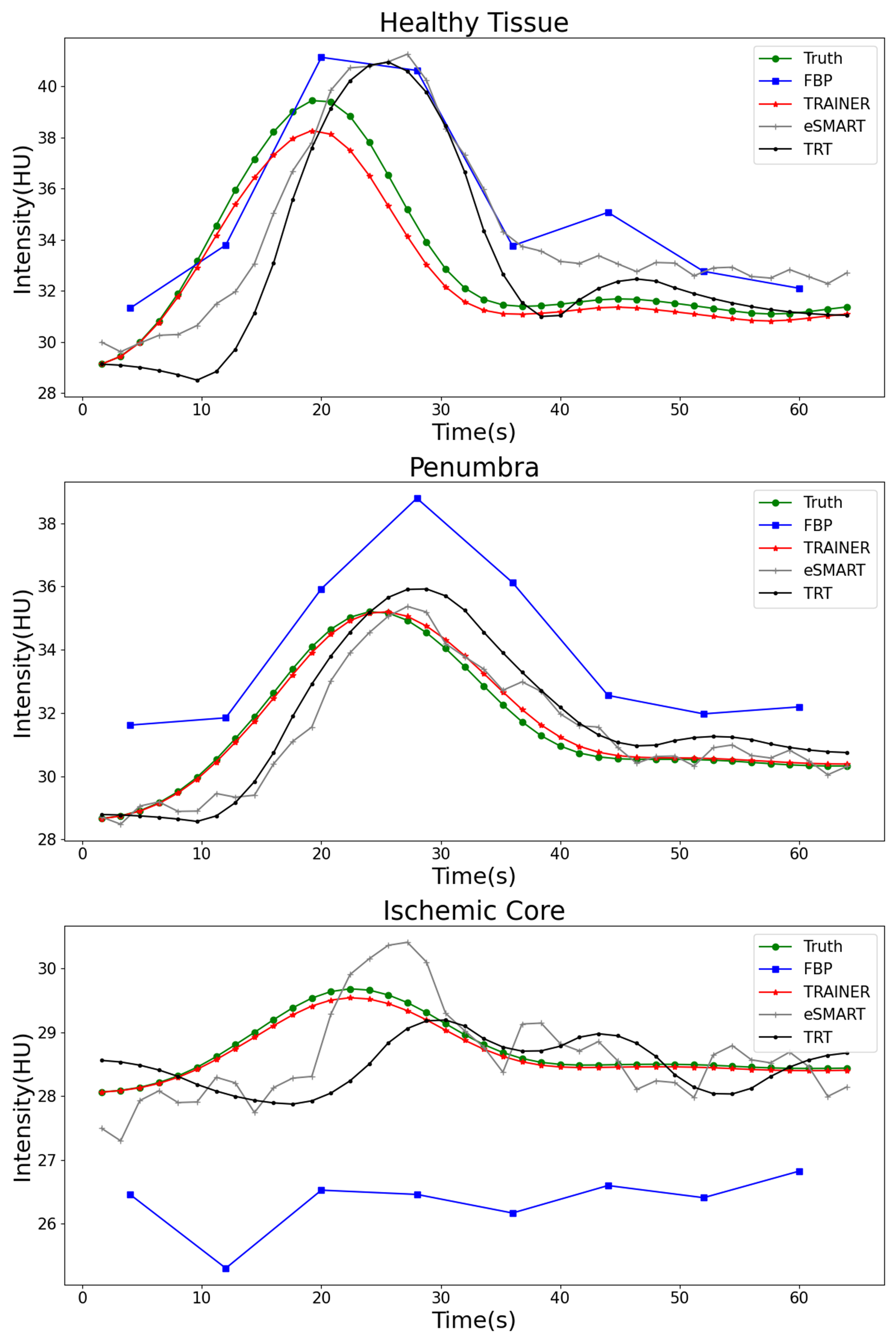}
		\caption{Comparison of time attenuation curves for different tissue types. The plots illustrate the time attenuation curves for healthy tissue (top), penumbra (middle), and ischemic core (bottom). The TACs generated by FBP (blue squares), TRT (black circles), eSMART (gray plus), and TRAINER (red stars) are compared with the ground truth (green circles).} %The results demonstrate that the curves produced by our method are consistently closer to the ground truth across all tissue types, confirming the superior performance of the proposed technique.
		\label{fig:tac-contrast}
	\end{figure}

	As shown in Fig.~\ref{fig:tac-contrast}, TRAINER consistently generates TACs closer to the ground truth (Truth) across three different tissue types: healthy tissue, penumbra, and ischemic core. Specifically, in the healthy tissue region, TACs obtained using TRAINER are nearly consistent with Truth, demonstrating a significant improvement on reconstruction accuracy over the baseline methods. In the penumbra region, although all methods show some fluctuations, TACs generated by TRAINER remains closer to Truth. In the ischemic core region, TRAINER yields smoother and more accurate TACs compared to the baseline methods, further highlighting the superior quantitative accuracy of TRAINER. These results further confirm the efficacy of TRAINER, demonstrating its superior performance in reconstructing TACs using data acquired by slow rotation speed systems.
	
	\subsection{Quantitative Assessment of Perfusion Parametric Images}
	
	The comparison of reconstruction error quantified by overall rRMSE and PSNR using baseline methods and TRAINER is shown in Tab.~\ref{tab:rmse} and Tab.~\ref{tab:psnr}. Compared to all baseline methods, TRAINER achieves the highest overall quantitative accuracy at all exposure levels. Additionally, we conducted a detailed statistical analysis on ten selected regions of interest (ROIs). The results demonstrated that in MTT for healthy tissue, penumbra, and ischemic core, the rRMSE metrics were significantly improved under four different exposure levels compared to the best performance of the baseline methods. Specifically, the rRMSE for healthy tissue decreased by an average of $60.56\%$, with a particularly notable decrease of $87.40\%$ in the penumbra region and a $72.45\%$ decrease in the ischemic core region. For CBF images in healthy tissue, penumbra, and ischemic core, the rRMSE decreased by averages of $42.22\%$, $15.26\%$, and $8.23\%$, respectively, under the four exposure levels, compared to the best performance of the baseline methods. These findings underscore the effectiveness of TRAINER in improving quantitative accuracy and image quality.
	
	\begin{table}[ht]  
		\centering  
		\caption{The comparison of reconstruction error quantified by overall rRMSE using baseline methods and TRAINER. Data were simulated at gantry rotation speed of $dt=8$ sec and four exposure levels.}  
		\label{tab:rmse}
		\setlength{\tabcolsep}{3pt} % 调整列间距
		\renewcommand{\arraystretch}{0.9} % 调整行间距
		\begin{tabular}{@{}l | l cccccc@{}} 
			\toprule  
			\textbf{} & \textbf{Method} & \textbf{$I_{0}=10^{6}$} & \textbf{$I_{0}=10^{5}$} & \textbf{$I_{0}=5\times 10^{4}$} & \textbf{$I_{0}=10^{4}$} \\   
			\midrule  
			\multirow{6}{*}{CBF} %& Fbp+sSVD & 19.28\% & 29.94\% & 35.49\% & 45.75\% \\ 
			%& Fbp+bSVD & 20.23\% & 31.14\% & 38.56\% & 48.24\% \\ 
			& Fbp+Tikh & 39.60\% & 39.61\% & 39.60\% & 39.68\% \\ 
			& Fbp+TTV  & 34.59\% & 34.28\% & 34.32\% & 34.54\% \\ 
			& TRT+TTV  & 12.97\% & 13.64\% & 14.62\% & 18.14\% \\ 
			& eSMART+TTV     & 13.49\% & 15.16\% & 15.74\% & 20.14\% \\ 
			& TRAINER & \textbf{4.94\%} & \textbf{6.11\%} & \textbf{7.24\%} & \textbf{10.46\%} \\  
			\midrule  
			\multirow{6}{*}{MTT} %& Fbp+sSVD & 47.82\% & 55.10\% & 54.50\% & 57.17\% \\ 
			%& Fbp+bSVD & 53.37\% & 48.92\% & 54.05\% & 54.35\% \\ 
			& Fbp+Tikh & 32.18\% & 32.15\% & 32.08\% & 31.69\% \\ 
			& Fbp+TTV  & 35.10\% & 34.96\% & 34.69\% & 33.84\% \\ 
			& TRT+TTV  & 18.64\% & 19.30\% & 19.34\% & 20.08\% \\ 
			& eSMART+TTV     & 23.57\% & 22.50\% & 22.56\% & 23.24\% \\ 
			& TRAINER & \textbf{3.60\%} & \textbf{7.08\%} & \textbf{5.63\%} & \textbf{8.31\%} \\    
			\bottomrule  
		\end{tabular}  
	\end{table}
	
	\begin{table}[ht]  
		\centering  
		\caption{The comparison of reconstruction error quantified by overall PSNR using baseline methods and TRAINER. Data were simulated at gantry rotation speed of $dt=8$ sec and four exposure levels.}  
		\label{tab:psnr}
		\setlength{\tabcolsep}{3pt} % 调整列间距
		\renewcommand{\arraystretch}{0.9} % 调整行间距
		\begin{tabular}{@{}l | l cccc@{}}   
			\toprule  
			\textbf{} & \textbf{Method} & \textbf{$I_{0}=10^{6}$} & \textbf{$I_{0}=10^{5}$} & \textbf{$I_{0}=5\times 10^{4}$} & \textbf{$I_{0}=10^{4}$} \\  
			\midrule  
			\multirow{6}{*}{CBF} %& Fbp+sSVD & 16.17 & 16.19 & 16.28 & 16.66 \\  
			%& Fbp+bSVD & 15.81 & 15.86 & 15.97 & 16.34 \\  
			& Fbp+Tikh & 15.05 & 15.05 & 15.05 & 15.03 \\  
			& Fbp+TTV  & 16.22 & 16.30 & 16.29 & 16.23 \\  
			& TRT+TTV  & 24.74 & 24.30 & 23.70 & 21.83 \\
			& eSMART+TTV      & 24.40 & 23.38 & 23.06 & 20.92 \\
			& TRAINER & \textbf{33.13} & \textbf{31.28} & \textbf{29.81} & \textbf{26.61} \\  
			\midrule  
			\multirow{6}{*}{MTT} %& Fbp+sSVD & 27.31 & 27.51 & 27.56 & 27.59 \\  
			%& Fbp+bSVD & 27.48 & 27.68 & 27.71 & 27.70 \\  
			& Fbp+Tikh & 26.13 & 26.13 & 26.15 & 26.26 \\   
			& Fbp+TTV  & 25.37 & 25.41 & 25.47 & 25.69 \\ 
			& TRT+TTV  & 30.87 & 30.57 & 30.55 & 30.22 \\
			& eSMART+TTV  & 28.83 & 29.24 & 29.21 & 28.95 \\
			& TRAINER & \textbf{45.16} & \textbf{39.27} & \textbf{41.26} & \textbf{37.89} \\  
			\bottomrule  
		\end{tabular}  
	\end{table}
	
\begin{figure}[htbp]
		\centering
		\includegraphics[width=1\linewidth]
		{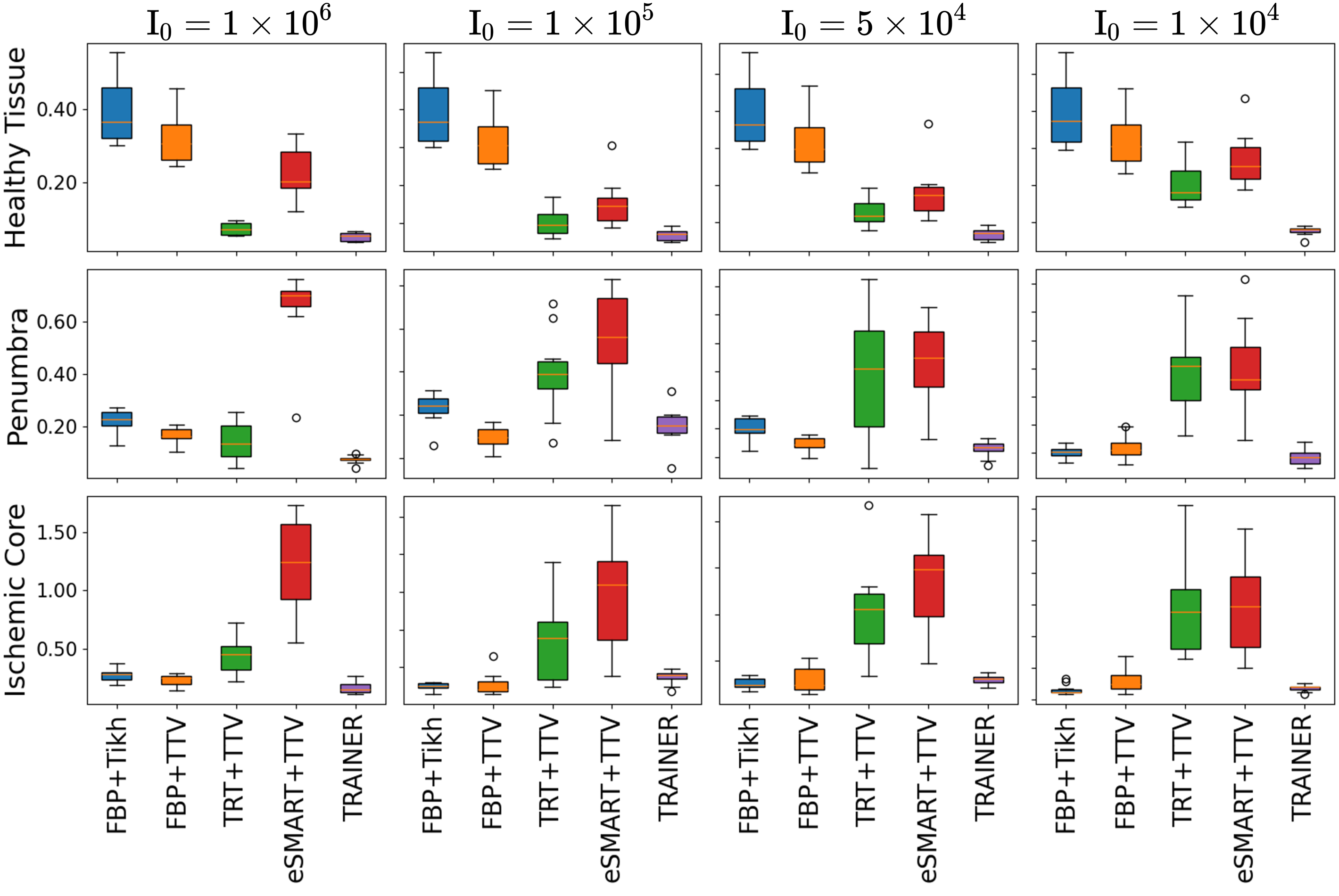}
		\caption{The comparison of reconstruction error of CBF quantified by tissue-specific rRMSE using baseline methods and TRAINER. Data were simulated at gantry rotation speed of $dt=8$ sec and four exposure levels.}
		\label{fig:cbf-rmse}
	\end{figure}

\begin{figure}[htbp]
		\centering
		\includegraphics[width=1\linewidth]
		{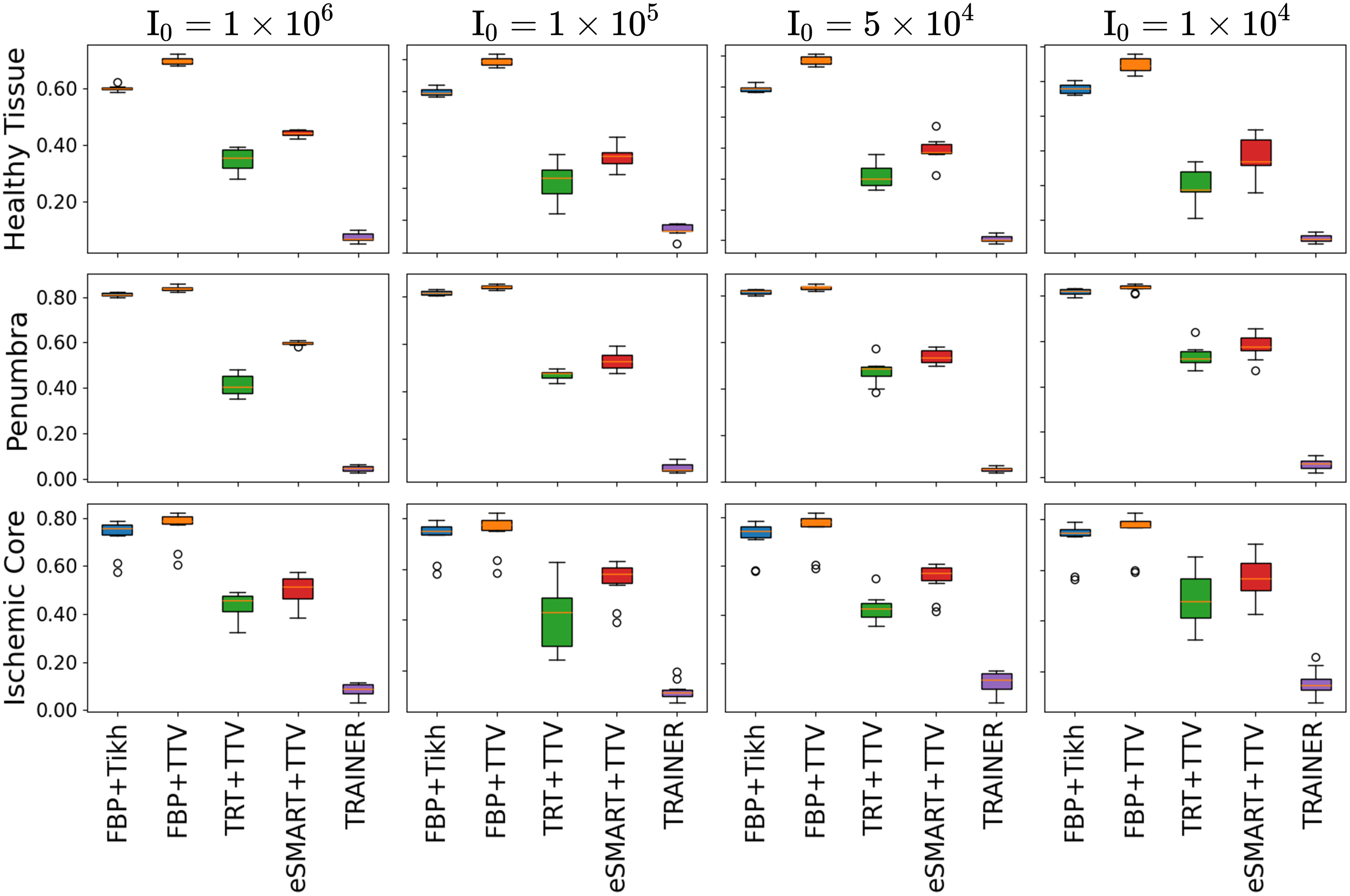}
		\caption{The comparison of reconstruction error of MTT quantified by tissue-specific rRMSE using baseline methods and TRAINER. Data were simulated at gantry rotation speed of $dt=8$ sec and four exposure levels.}
		\label{fig:mtt-rmse}
	\end{figure}

	The comparison of reconstruction error quantified by rRMSE for each tissue type (healthy tissue, penumbra, ischemic core) using baseline methods and TRAINER was shown in Fig.~\ref{fig:cbf-rmse} and Fig.~\ref{fig:mtt-rmse}. Data were simulated at gantry rotation speed of $dt=8$ sec and four exposure levels. Compared to the baseline methods, TRAINER improves the quantitative accuracy of perfusion parametric images for all types of tissues. rRMSE values were calculated by carefully selecting ten regions of interest.
	
	%%%%%%%%%%%%%%%%%%%%%%%%%%%%%%%%%%%%%%%%%%%%%%%%%%%%%%%%%%%%%%%%%%%%%%%%%%%%%%%%%%%%%%%%%%%%%%%%%%%%%
	
	\section{Discussion and Conclusion}
	
	\subsection{Potential Limitations and Future Work}
	
	This work has several potential limitations which need further investigations in future studies. 
	
	First, the performance dependence of TRAINER on optimization algorithms has not been investigated yet. Currently, the widely used Adam was applied to solve Eq.~(\ref{eq:subproblem-1}) and the stochastic gradient descent was applied to solve Eq.~(\ref{eq:subproblem-2}). Apparently other optimization algorithms can also be used as a candidate optimizer for these two problems. It is necessary to compare different optimization algorithms to solve Eq.~(\ref{eq:subproblem-1}) and Eq.~(\ref{eq:subproblem-2}) in terms of stability and convergence speed. 
	
	Second, the subject-specific generative model in Eq.~(\ref{eq:generative-model}) is conditioned by its input, $\mathbf{Z}$. Currently, an empirical selection of $\mathbf{Z}$ has been employed to generate results shown in the paper. It is very likely that the quantitative accuracy and image quality of the reconstructed perfusion parametric images depend on the selection of $\mathbf{Z}$. In our future studies, we will investigate the performance dependence on different selections such as time-resolved FBP images. In addition, the condition $\mathbf{Z}$ may not necessarily be treated as a constant, instead, it can be treated as an argument to be optimized under the ADMM iterative framework. In our future studies, we will also explore the technical feasibility of optimizing the condition and investigate the potential benefits of this novel strategy. 
	
	\subsection{Conclusion}
	
	In this work, we demonstrated that using TRAINER, cerebral perfusion parametric images can be accurately obtained via training a subject-specific conditional generative model under the constraint of the subject's own measured data. Results shown in the paper demonstrated that the two technical challenges, i.e. poor temporal resolution of C-arm CBCT and inaccurate perfusion parametric estimation methods, have been simultaneously addressed using the developed TRAINER technique. 
	
	%%%%%%%%%%%%%%%%%%%%%%%%%%%%%%%%%%%%%%%%%%%%%%%%%%%%%%%%%%%%%%%%%%%%%%%%%%%%%%%%%%%%%%%%%%%%%%%%%%%%%
	
	\IEEEtriggeratref{70}
	
	\bibliographystyle{IEEEtran}
	
	\bibliography{cerebral}

\begin{thebibliography}{10}
\providecommand{\url}[1]{#1}
\csname url@samestyle\endcsname
\providecommand{\newblock}{\relax}
\providecommand{\bibinfo}[2]{#2}
\providecommand{\BIBentrySTDinterwordspacing}{\spaceskip=0pt\relax}
\providecommand{\BIBentryALTinterwordstretchfactor}{4}
\providecommand{\BIBentryALTinterwordspacing}{\spaceskip=\fontdimen2\font plus
\BIBentryALTinterwordstretchfactor\fontdimen3\font minus
  \fontdimen4\font\relax}
\providecommand{\BIBforeignlanguage}[2]{{%
\expandafter\ifx\csname l@#1\endcsname\relax
\typeout{** WARNING: IEEEtran.bst: No hyphenation pattern has been}%
\typeout{** loaded for the language `#1'. Using the pattern for}%
\typeout{** the default language instead.}%
\else
\language=\csname l@#1\endcsname
\fi
#2}}
\providecommand{\BIBdecl}{\relax}
\BIBdecl

\bibitem{RN9}
D.~G. Nabavi, A.~Cenic, R.~A. Craen, A.~W. Gelb, J.~D. Bennett, R.~Kozak, and
  T.-Y. Lee, ``Ct assessment of cerebral perfusion: experimental validation and
  initial clinical experience,'' \emph{Radiology}, vol. 213, no.~1, pp.
  141--149, 1999.

\bibitem{RN126}
R.~G. González, ``Clinical mri of acute ischemic stroke,'' \emph{Journal of
  Magnetic Resonance Imaging}, vol.~36, no.~2, pp. 259--271, 2012.

\bibitem{RN65}
P.~Khatri, S.~D. Yeatts, M.~Mazighi, J.~P. Broderick, D.~S. Liebeskind, A.~M.
  Demchuk, P.~Amarenco, J.~Carrozzella, J.~Spilker, and L.~D. Foster, ``Time to
  angiographic reperfusion and clinical outcome after acute ischaemic stroke:
  an analysis of data from the interventional management of stroke (ims iii)
  phase 3 trial,'' \emph{The Lancet Neurology}, vol.~13, no.~6, pp. 567--574,
  2014.

\bibitem{niu2016c}
K.~Niu, P.~Yang, Y.~Wu, T.~Struffert, A.~Doerfler, S.~Schafer, K.~Royalty,
  C.~Strother, and G.-H. Chen, ``C-arm conebeam ct perfusion imaging in the
  angiographic suite: a comparison with multidetector ct perfusion imaging,''
  \emph{American Journal of Neuroradiology}, vol.~37, no.~7, pp. 1303--1309,
  2016.

\bibitem{yang2015time}
P.~Yang, K.~Niu, Y.~Wu, T.~Struffert, A.~Dorfler, S.~Schafer, K.~Royalty,
  C.~Strother, and G.-H. Chen, ``Time-resolved c-arm computed tomographic
  angiography derived from computed tomographic perfusion acquisition: new
  capability for one-stop-shop acute ischemic stroke treatment in the
  angiosuite,'' \emph{Stroke}, vol.~46, no.~12, pp. 3383--3389, 2015.

\bibitem{li2019enhanced}
Y.~Li, J.~W. Garrett, K.~Li, C.~Strother, and G.-H. Chen, ``An enhanced
  smart-recon algorithm for time-resolved c-arm cone-beam ct imaging,''
  \emph{IEEE transactions on medical imaging}, vol.~39, no.~6, pp. 1894--1905,
  2019.

\bibitem{neukirchen2010iterative}
C.~Neukirchen, M.~Giordano, and S.~Wiesner, ``An iterative method for
  tomographic x-ray perfusion estimation in a decomposition model-based
  approach,'' \emph{Medical physics}, vol.~37, no.~12, pp. 6125--6141, 2010.

\bibitem{wagner2013model}
M.~Wagner, Y.~Deuerling-Zheng, M.~M{\"o}hlenbruch, M.~Bendszus, J.~Boese, and
  S.~Heiland, ``A model based algorithm for perfusion estimation in
  interventional c-arm ct systems,'' \emph{Medical physics}, vol.~40, no.~3, p.
  031916, 2013.

\bibitem{manhart2013dynamic}
M.~T. Manhart, M.~Kowarschik, A.~Fieselmann, Y.~Deuerling-Zheng, K.~Royalty,
  A.~K. Maier, and J.~Hornegger, ``Dynamic iterative reconstruction for
  interventional 4-d c-arm ct perfusion imaging,'' \emph{IEEE transactions on
  medical imaging}, vol.~32, no.~7, pp. 1336--1348, 2013.

\bibitem{tang2013novel}
J.~Tang, M.~Xu, K.~Niu, K.~Royalty, K.~Pulfer, C.~Strother, and G.-H. Chen, ``A
  novel temporal recovery technique to enable cone beam ct perfusion imaging
  using an interventional c-arm system,'' in \emph{Medical Imaging 2013:
  Physics of Medical Imaging}, vol. 8668.\hskip 1em plus 0.5em minus
  0.4em\relax International Society for Optics and Photonics, 2013, p. 86681A.

\bibitem{van2016local}
V.~Van~Nieuwenhove, G.~Van~Eyndhoven, K.~J. Batenburg, N.~Buls,
  J.~Vandemeulebroucke, J.~De~Beenhouwer, and J.~Sijbers, ``Local attenuation
  curve optimization framework for high quality perfusion maps in low-dose
  cerebral perfusion ct,'' \emph{Medical physics}, vol.~43, no.~12, pp.
  6429--6438, 2016.

\bibitem{chen2015synchronized}
G.-H. Chen and Y.~Li, ``Synchronized multiartifact reduction with tomographic
  reconstruction (smart-recon): A statistical model based iterative image
  reconstruction method to eliminate limited-view artifacts and to mitigate the
  temporal-average artifacts in time-resolved ct,'' \emph{Medical physics},
  vol.~42, no.~8, pp. 4698--4707, 2015.

\bibitem{li2018time}
Y.~Li, J.~W. Garrett, K.~Li, Y.~Wu, K.~Johnson, S.~Schafer, C.~Strother, and
  G.-H. Chen, ``Time-resolved c-arm cone beam ct angiography (tr-cbcta) imaging
  from a single short-scan c-arm cone beam ct acquisition with intra-arterial
  contrast injection,'' \emph{Physics in Medicine \& Biology}, vol.~63, no.~7,
  p. 075001, 2018.

\bibitem{Li2023airport}
Y.~Li, J.~Feng, J.~Xiang, Z.~Li, and D.~Liang, ``Airport: A data consistency
  constrained deep temporal extrapolation method to improve temporal resolution
  in contrast enhanced ct imaging,'' \emph{IEEE transactions on medical
  imaging}, 2023.

\bibitem{RN74}
R.~Fang, S.~Zhang, T.~Chen, and P.~C. Sanelli, ``Robust low-dose ct perfusion
  deconvolution via tensor total-variation regularization,'' \emph{IEEE
  transactions on medical imaging}, vol.~34, no.~7, pp. 1533--1548, 2015.

\bibitem{RN73}
C.~Frindel, M.~C. Robini, and D.~Rousseau, ``A 3-d spatio-temporal
  deconvolution approach for mr perfusion in the brain,'' \emph{Medical image
  analysis}, vol.~18, no.~1, pp. 144--160, 2014.

\bibitem{RN72}
T.~Boutelier, K.~Kudo, F.~Pautot, and M.~Sasaki, ``Bayesian hemodynamic
  parameter estimation by bolus tracking perfusion weighted imaging,''
  \emph{IEEE transactions on medical imaging}, vol.~31, no.~7, pp. 1381--1395,
  2012.

\bibitem{RN71}
R.~Fang, T.~Chen, and P.~C. Sanelli, ``Towards robust deconvolution of low-dose
  perfusion ct: Sparse perfusion deconvolution using online dictionary
  learning,'' \emph{Medical image analysis}, vol.~17, no.~4, pp. 417--428,
  2013.

\bibitem{RN70}
L.~He, B.~Orten, S.~Do, W.~C. Karl, A.~Kambadakone, D.~V. Sahani, and H.~Pien,
  ``A spatio-temporal deconvolution method to improve perfusion ct
  quantification,'' \emph{IEEE Transactions on Medical Imaging}, vol.~29,
  no.~5, pp. 1182--1191, 2010.

\bibitem{RN69}
D.~Zeng, X.~Zhang, Z.~Bian, J.~Huang, H.~Zhang, L.~Lu, W.~Lyu, J.~Zhang,
  Q.~Feng, and W.~Chen, ``Cerebral perfusion computed tomography deconvolution
  via structure tensor total variation regularization,'' \emph{Medical
  physics}, vol.~43, no.~5, pp. 2091--2107, 2016.

\bibitem{RN68}
K.~Mouridsen, K.~Friston, N.~Hjort, L.~Gyldensted, L.~Østergaard, and
  S.~Kiebel, ``Bayesian estimation of cerebral perfusion using a physiological
  model of microvasculature,'' \emph{Neuroimage}, vol.~33, no.~2, pp. 570--579,
  2006.

\bibitem{RN55}
C.~H. Cremers, J.~W. Dankbaar, M.~D. Vergouwen, P.~C. Vos, E.~Bennink, G.~J.
  Rinkel, B.~K. Velthuis, and I.~C. van~der Schaaf, ``Different ct perfusion
  algorithms in the detection of delayed cerebral ischemia after aneurysmal
  subarachnoid hemorrhage,'' \emph{Neuroradiology}, vol.~57, no.~5, pp.
  469--474, 2015.

\bibitem{RN41}
R.~Mangla, S.~Ekhom, B.~S. Jahromi, J.~Almast, M.~Mangla, and P.-L. Westesson,
  ``Ct perfusion in acute stroke: know the mimics, potential pitfalls,
  artifacts, and technical errors,'' \emph{Emergency radiology}, vol.~21,
  no.~1, pp. 49--65, 2014.

\bibitem{miles2007multidetector}
K.~Miles, J.~D. Eastwood, and M.~K{\"o}nig, \emph{Multidetector Computed
  Tomography in Cerebrovascular Disease: CT Perfusion Imaging}.\hskip 1em plus
  0.5em minus 0.4em\relax Boca Raton, FL, USA: CRC Press, 2007.

\bibitem{RN125}
A.~Fieselmann, M.~Kowarschik, A.~Ganguly, J.~Hornegger, and R.~Fahrig,
  ``Deconvolution-based ct and mr brain perfusion measurement: theoretical
  model revisited and practical implementation details,'' \emph{International
  Journal of Biomedical Imaging}, vol. 2011, 2011.

\bibitem{RN124}
A.~Konstas, G.~Goldmakher, T.-Y. Lee, and M.~Lev, ``Theoretic basis and
  technical implementations of ct perfusion in acute ischemic stroke, part 1:
  theoretic basis,'' \emph{American Journal of Neuroradiology}, vol.~30, no.~4,
  pp. 662--668, 2009.

\bibitem{RN123}
------, ``Theoretic basis and technical implementations of ct perfusion in
  acute ischemic stroke, part 2: technical implementations,'' \emph{American
  Journal of Neuroradiology}, vol.~30, no.~5, pp. 885--892, 2009.

\bibitem{RN46}
R.~M. Ferreira, M.~H. Lev, G.~V. Goldmakher, S.~Kamalian, P.~W. Schaefer, K.~L.
  Furie, R.~G. Gonzalez, and P.~C. Sanelli, ``Arterial input function placement
  for accurate ct perfusion map construction in acute stroke,'' \emph{AJR.
  American journal of roentgenology}, vol. 194, no.~5, p. 1330, 2010.

\bibitem{ulyanov2018deep}
D.~Ulyanov, A.~Vedaldi, and V.~Lempitsky, ``Deep image prior,'' in
  \emph{Proceedings of the IEEE conference on computer vision and pattern
  recognition}, 2018, pp. 9446--9454.

\bibitem{kingma2014adam}
D.~P. Kingma and J.~Ba, ``Adam: A method for stochastic optimization,''
  \emph{arXiv preprint arXiv:1412.6980}, 2014.

\bibitem{dmitry2020deep}
U.~Dmitry, A.~Vedaldi, and L.~Victor, ``Deep image prior,'' \emph{International
  Journal of Computer Vision}, vol. 128, no.~7, pp. 1867--1888, 2020.

\bibitem{he2015delving}
K.~He, X.~Zhang, S.~Ren, and J.~Sun, ``Delving deep into rectifiers: Surpassing
  human-level performance on imagenet classification,'' in \emph{Proceedings of
  the IEEE international conference on computer vision}, 2015, pp. 1026--1034.

\bibitem{RN183}
A.~Aichert, M.~T. Manhart, B.~K. Navalpakkam, R.~Grimm, J.~Hutter, A.~Maier,
  J.~Hornegger, and A.~Doerfler, ``A realistic digital phantom for perfusion
  c-arm ct based on mri data,'' in \emph{2013 IEEE Nuclear Science Symposium
  and Medical Imaging Conference (2013 NSS/MIC)}.\hskip 1em plus 0.5em minus
  0.4em\relax IEEE, 2013, Conference Proceedings, pp. 1--2.

\bibitem{fieselmann2011deconvolution}
A.~Fieselmann, M.~Kowarschik, A.~Ganguly, J.~Hornegger, and R.~Fahrig,
  ``Deconvolution-based ct and mr brain perfusion measurement: theoretical
  model revisited and practical implementation details,'' \emph{International
  Journal of Biomedical Imaging}, vol. 2011, no.~1, p. 467563, 2011.

\bibitem{fang2015robust}
R.~Fang, S.~Zhang, T.~Chen, and P.~C. Sanelli, ``Robust low-dose ct perfusion
  deconvolution via tensor total-variation regularization,'' \emph{IEEE
  transactions on medical imaging}, vol.~34, no.~7, pp. 1533--1548, 2015.

\bibitem{kudo2010differences}
K.~Kudo, M.~Sasaki, K.~Yamada, S.~Momoshima, H.~Utsunomiya, H.~Shirato, and
  K.~Ogasawara, ``Differences in ct perfusion maps generated by different
  commercial software: quantitative analysis by using identical source data of
  acute stroke patients,'' \emph{Radiology}, vol. 254, no.~1, pp. 200--209,
  2010.

\end{thebibliography}
	
\end{document}